# LSTM-Based Net Load Forecasting for Wind and Solar Power-Equipped Microgrids

Jesus Silva-Rodriguez
Department of Electrical and
Computer Engineering
University of Houston
Houston, TX, USA
jasilvarodriguez@uh.edu

Elias Raffoul
Department of Electrical and
Computer Engineering
University of Houston
Houston, TX, USA
ejraffoul@uh.edu

Xingpeng Li
Department of Electrical and
Computer Engineering
University of Houston
Houston, TX, USA
xli82@uh.edu

*Abstract*—The rising integration of variable renewable energy sources (RES), like solar and wind power, introduces considerable uncertainty in grid operations and energy management. Effective forecasting models are essential for grid operators to anticipate the net load—the difference between consumer electrical demand and renewable power generation. This paper proposes a deep learning (DL) model based on long short-term memory (LSTM) networks for net load forecasting in renewable-based microgrids, considering both solar and wind power. The model's architecture is detailed, and its performance is evaluated using a residential microgrid test case based on a typical meteorological year (TMY) dataset. The results demonstrate the effectiveness of the proposed LSTM-based DL model in predicting the net load, showcasing its potential for enhancing energy management in renewable-based microgrids.

*Index Terms*-- Long Short-Term Memory, Microgrid, Net Load Forecasting, Recurrent Neural Network, Renewable Energy.

## I. Introduction

The increased penetration of variable renewable energy sources (RES) such as solar and wind power creates substantial levels of uncertainty on gird operation and energy management. Because RESs are essentially considered as "free clean energy" sources, given that they technically require zero fuel, then it is desirable to use as much of their generation as it is available, and meet the remaining demand with conventional controllable generators such as fossil fuel-based thermal generators. For grid operators to understand how to perform energy management to meet the imbalance between consumer electrical demand and RES generation, known as the net load, adequate forecasting models must be implemented to predict good approximations of what this net load will be in the near future.

The concept of load forecasting has already been explored in literature using several different methods that range from statistical methods to machine learning (ML) methods [1]-[6]. Moreover, net load forecasting (NLF) has also been explored to some extent recently, implementing different state-of-the-art ML models and algorithms to predict the combination of consumer demand and renewable power generation [7]-[10].

ML algorithms, especially deep learning (DL), are models that are capable of exceptionally accurate predictions. Several of them perform particularly well on time series predictions, capable of learning even some of the most complicated non-linear relations between certain parameters [11]. Hence there is a strong appeal for these models to be used for load forecasting given the highly variable nature of load patterns which depend on many different factors.

DL is used in the form of an unshared convolutional neural network in [2], which proved to perform better than fully connected networks in terms of number of parameters and vulnerability to overfitting. Moreover, a pooling-based deep recurrent neural network (RNN) is proposed in [3], which is able to directly learn uncertainties for household load forecasting using historical data of not only the consumer, but also its neighbors in order to allow the model to also learn correlations and interactions in residential areas. Additionally, [3] also implements a long-short term memory (LSTM) unit to improve its household load forecast.

An LSTM model combined with the convolutional neural network (CNN) concept is implemented in [4] in a hybrid LSTM-CNN approach that can forecast short-term load demand. Moreover, the performance of LSTM is further demonstrated in [5] that compared with other DL models such as temporal CNNs and support vector regression, LSTM is proven to outperform all of them for the task of load forecasting. LSTM is a powerful RNN model that was specially designed to improve stability and performance over deep-RNN models when learning long-term dependencies, and as such, is an excellent model to handle scenarios that involve time series predictions [12]. Therefore, it is a good candidate for NLF as well.

The Bayesian neural network (BNN) model, a technique incorporating stochasticity to the model's learning process which is useful to handle the inherent uncertainty of RES, also seems to be a good-performing model. For example, [8] proposes a methodology for short term NLF that uses a BNN model. This BNN model is used in [9] along with LSTM, combining the Bayesian statistical theory and deep LSTM networks to achieve more accurate forecasts with less data than other conventional DL models.

Additionally, the majority of research in current literature focuses on solar generation as the only RES present in the net load, and the inclusion of wind power seems to be scarce. Moreover, deep neural networks are used in [10] to forecast a net load that is composed of consumer demand, solar and wind power generation. However, the temporal correlations that may exist in demand, solar and wind power data are not explored. Therefore, this paper proposes a DL model based on LSTM to create NLF profiles for renewable-based microgrids that incorporate both solar and wind power.

In addressing the intricacies of RES integration, it is essential to understand the operational dynamics of a microgrid. The latter is a compact electrical network designed to generate and distribute power locally. It integrates diverse energy sources like solar, wind, or fuel, along with storage solutions such as batteries or electric vehicles. Operating in two modes, a microgrid can seamlessly connect to the main grid, exchanging power, or function independently in islanded mode, disconnected from the main grid [13]-[17]. Fig. 1 displays a diagram of a typical architecture of a microgrid network.

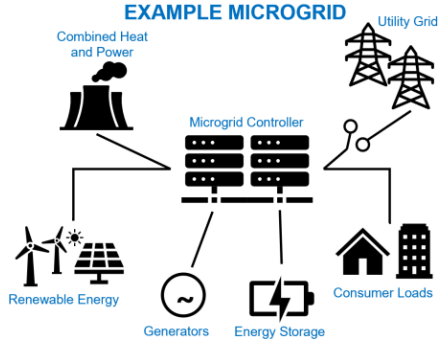

Fig. 1: Example microgrid network.

Yearly data for a typical meteorological year (TMY) is used to obtain a test case for a microgrid serving a residential area, and short-term NLF is carried out and evaluated against the actual data to determine the effectiveness and prediction accuracy of the proposed model.

The rest of the paper is organized as follows. Section II covers the methodology used to derive the solar and wind power data for the microgrid test case, based on the TMY data on ambient temperature, wind speed, and solar irradiance [19], [20], as well as the DL model architecture implementing LSTM units. Section III describes the TMY dataset used, presents the residential microgrid test case, and explains how the LSTM-based DL model is trained. Section IV presents the prediction results along with the scoring metrics to evaluate the quality of the prediction. Lastly, Section V concludes the paper.

## II. METHODOLOGY

The purpose and the idea behind the proposed model are that with given expected meteorological parameter values for ambient temperature, wind speed, and solar irradiance, the model can process this information along with the time of the year and the day, and use that to predict the net load of the microgrid. Since only meteorological data and average residential unit power demand data are available, the net wind and solar power generation data must be derived based on the meteorological parameters available, and once ready, it can be combined with power demand to create the net load label data to construct and train the DL model.

### A. Wind Power

Wind power generation from wind turbines is primarily dependent on the speed of the wind that circulates through the turbine's blades. The available power that can be generated is then related to wind speed by the mathematical function presented in (1), where $v$ is the wind speed, $\rho_0$ is the air density, $\eta_w$ is the wind turbine efficiency, and $A$ is the turbine blade swept area, given by (2) where $D$ represents the diameter, or the blade's length [14].

$$P^{WT} = \tfrac{1}{2}\rho_0 A v^3 \eta_w \quad (1)$$

$$A = \tfrac{\pi}{4} D^2 \quad (2)$$

However, the actual output power of wind turbines is limited by their operating rated wind speed. Normally, wind turbines have a specific cut-in speed $v_{ci}$, cut-out speed $v_{co}$, and rated speed $v_r$ that along with the current speed passing through the turbine, determine the actual output of the wind turbine [19], [21]. The wind power from a single turbine is then determined with (3).

$$P^{WT}(v) = \begin{cases} \tfrac{1}{2}\rho_0 A v^3 \eta_w, & v_{ci} \le v < v_r \\ P_r, & v_r \le v \le v_{co} \\ 0, & v < v_{ci} \text{ or } v > v_{co} \end{cases} \quad (3)$$

Using (3) then wind power output data can be calculated for every time-series interval, depending on the TMY data for wind speed [19].

### B. Solar Power

Solar power generated by photovoltaic (PV) systems is a function of the incoming solar irradiance on the surface of the PV panel, and is given by (4), where $\eta_{pv}$ is the PV panel efficiency, $S_{pv}$ is the surface area of the panel, and $I_C$ is the total solar irradiance collected by the surface of the panel [21]-[23]. However, normally the efficiency of PV systems is not a constant parameter, and instead it is affected by the PV cell temperature, which in turn affects power output. Normally in literature this is expressed as a linear regression based on parameters that are given in terms of standard testing conditions (STC) as reference points, and a temperature relation which depends on the material of the solar panel, given by (5) where $I_{C,ref}$ and $T_{ref}$ are the reference PV collector irradiance and cell temperature at STC, which give the rated PV power output $P_{pv,ref}$, and $\gamma_{ref}$ is the temperature relation coefficient, and $T_{cell}$ is the PV cell temperature [19], [20], [21].

$$P_{pv} = \eta_{pv} S_{pv} I_C \quad (4)$$

$$P_{pv} = \tfrac{I_C}{I_{C,ref}} P_{pv,ref} \cdot [1 - \gamma_{ref}(T_{cell} - T_{ref})] \quad (5)$$

Equation (5) effectively captures the cell temperature relation with output power. However, cell temperature must be determined as well, which will depend on the ambient conditions of the surroundings. Energy transfer between different media occurs as a result of a temperature difference, and it generally occurs via three heat transfer paths: heat absorption, heat radiation, and heat convection [24]-[25].

In a closed system, heat balance occurs, where the three different heat transfer paths balance each other out. Therefore, for a single time interval assuming steady-state conditions, the heat balance for a PV system would be given as in (6), where $q_s$, $q_r$, and $q_c$ are the heat absorption, heat radiation, and heat convection rates, respectively [25].

$$q_s + q_c + q_r - P_{pv} = 0 \quad (6)$$

For the PV panel case, heat absorption is defined by (7), where $\alpha_{cell}$ represents the PV cell absorptivity, heat radiation is defined by (8), where $\sigma$ is the Stefan-Boltzmann constant ($5.669 \times 10^{-8}$ W/m$^2$K$^4$), $\varepsilon_a$ the emissivity of the surroundings, $T_a$ the ambient temperature, and $\varepsilon_{cell}$ the PV cell emissivity, and heat convection is defined by (9), where $h_c$ is a convective heat transfer coefficient [25].

$$q_s = \alpha_{cell} I_C S_{pv} \quad (7)$$

$$q_r = S_{pv}\sigma[\varepsilon_a T_a^4 - \varepsilon_{cell} T_{cell}^4] \quad (8)$$

$$q_c = -h_c S_{pv}(T_{cell} - T_a) \quad (9)$$

The heat transfer coefficient $h_c$ is defined by (10) as the sum of two other coefficients known as the free convection $h_{c,free}$ and the forced convection $h_{c,forced}$ coefficients, which in turn depend on the ambient conditions, especially wind speed and temperature. For PV panels these two are defined by (11) and (12), respectively [24].

$$h_c = h_{c,free} + h_{c,forced} \quad (10)$$

$$h_{c,free} = \frac{0.1 k_0}{L}\left(\frac{g\rho_0 \beta_0 C_{p0}}{\mu_0 k_0}\right)^{\frac{1}{3}}(T_{cell} - T_a)^{\frac{1}{3}} \quad (11)$$

$$h_{c,forced} = \begin{cases} \frac{0.664 k_0}{L}\left(\frac{\rho_0 v}{L}\right)^{\frac{1}{2}}, v < 3.3037\ m/s \\ \frac{0.037 k_0^{\frac{2}{3}}}{L^{\frac{3}{10}}\mu_0^{\frac{7}{15}}}(\rho_0 v)^{\frac{4}{5}}, v > 3.3037\ m/s \end{cases} \quad (12)$$

Combining (5), (7)-(12) into (6) creates a function for cell temperature in terms of PV collector irradiance, ambient temperature, and wind speed, all of which can be obtained from the meteorological TMY data [19], [20]. Using this function then PV cell temperature can be determined for every time-series interval and in turn be used in (5) to calculate the solar power output data.

### C. LSTM-based Deep Learning Model

LSTM is one of the most advanced DL networks to process temporal sequences, which makes it an ideal choice for microgrid NLF. The LSTM architecture consists of a set of recurrently connected subnetworks known as memory blocks, which maintain their state over time and regulate the flow of information through non-linear gating units [12]. Fig. 2 displays a diagram of a typical architecture of an LSTM block. Following the flow of information through the unit, the LSTM block determines what information to remove from its previous state, and what information to continue passing along as part of the output [26].

Using this block from Fig. 2 as the base for the NLF, the proposed model of this paper implements multiple LSTM layers along with a couple of dense layers to obtain the best prediction result possible. Through exhaustive attempts which considered multiple combinations of hyperparameters such as the number of LSTM and dense layers and units, the best combination was determined to be three LSTM layers which take and process the input data to extract and learn temporal relations that exist between input data and the net load labels, which are composed of the combination of solar and wind power determined with (1)-(3) and (5)-(12), respectively, and the load demand data obtained from a TMY dataset from [27]. Additionally, all LSTM layers apply L2 regularization with a lambda value of 0.001 for the kernel and recurrent weights, dropout rates of 0.4 to prevent overfitting, and utilize batch normalization to stabilize the learning process and accelerate training. After these three LSTM layers, a dense layer with a linear activation follows to determine more correlations between the input data and the labels and preserve the potentially negative values that can occur on net load in cases when renewable generation is higher than load demand. Lastly a final output layer is introduced to condense all the information into a prediction of the net load. The diagram in Fig. 3 summarizes graphically the architecture of the proposed model for the NLF of renewable-based microgrids.

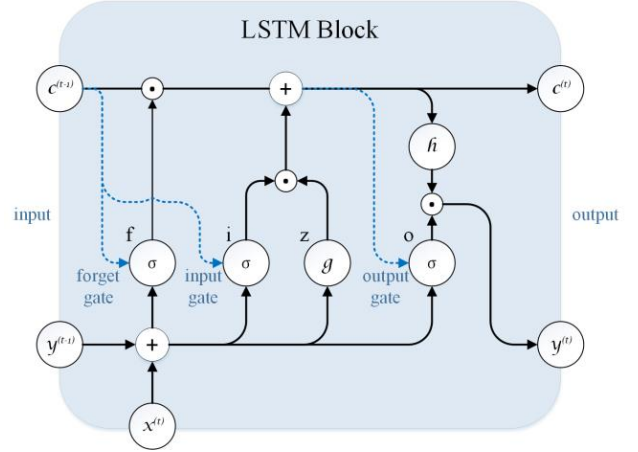

Fig. 2: Architecture of a typical LSTM block.

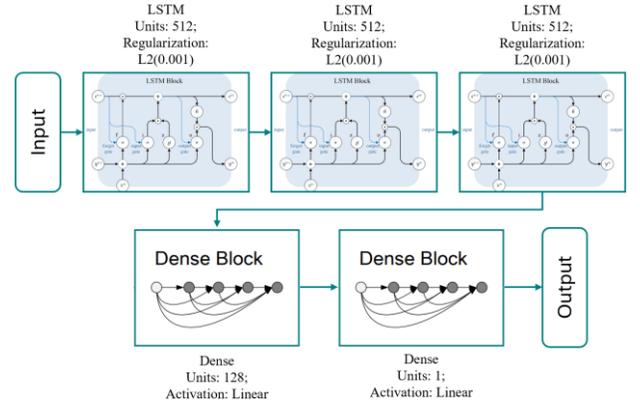

Fig. 3: LSTM-based model architecture for microgrid net load forecasting.

### D. Direct and Indirect NLF Approaches

The LSTM-based model is then used to predict the net load of this microgrid in two different ways: a direct approach shown in Fig. 4, which takes the input data and directly trains the model with the net load derived data as the label, and an indirect approach, which takes the input data and trains the model in three different instances to predict solar power, wind power, and load demand separately, to later combine them to calculate the net load as in Fig. 5 [10]. A comparison will then be carried out to determine the most accurate method.

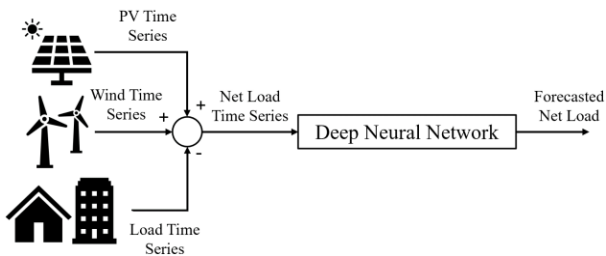

Fig. 4: Direct approach net load prediction.

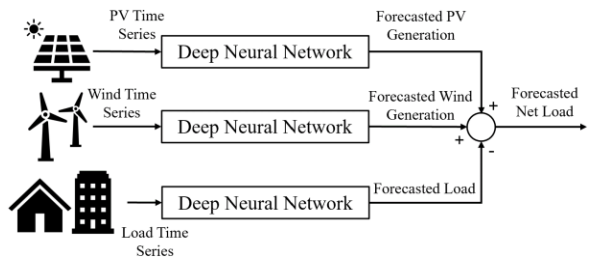

Fig. 5: Indirect approach net load prediction.

## III. CASE STUDIES

### A. Data Description

The meteorological data from [19] includes hourly ambient temperature and wind speed data for multiple locations in the state of Texas for the year of 2019. Similarly, the data of [20] includes global horizontal irradiance (GHI), direct horizontal irradiance (DHI), and direct normal irradiance (DNI) for a typical year for multiple locations in the United States, including Texas. Lastly the data from [27] includes average load for residential units for a typical year for the same locations. Load demand is part of the net load that wishes to be forecasted, and that would be part of the data labels for training of the LSTM-based model. However, net load is calculated as the difference between load demand and the combined output power of the solar and wind power systems. Thus, the data from [19]-[20] for wind speed, temperature and irradiance is used along with the theory from Section II to determine the solar and wind power output of the microgrid.

The case scenario used to test the proposed LSTM-based model for microgrid NLF corresponds to a synthetic test case of a residential microgrid assumed to be located in the Houston area, composed of 60 residential units, 100 Tesla T430S PV modules [28], and 3 small-scale wind turbines [29].

### B. Model Training

In both the direct and indirect methodologies, the input dataset remains consistent, maintaining a size of (8760, 5). Here, 8760 signifies the product of 24 hours and 365 days, with the five columns representing the day number, hour of the day, temperature (K), wind speed (m/s), and collector irradiance (W/m$^2$). On the output side, the dataset will be of size (8760, 1). In the direct approach, it corresponds to the net load to be predicted, while for the indirect approach, load demand, wind power, and solar power will be predicted separately.

To enhance the training process, the input dataset undergoes normalization, promoting better convergence of the training model. Additionally, the input data is partitioned into three sets. The training dataset constitutes 80% of the values, resulting in a size of (7008, 5). Two datasets of 10% each, named the validation dataset and testing dataset with sizes (876, 5), are also created. This trisection serves specific functions to aid in optimal training convergence. The training dataset facilitates model learning, the validation dataset assists in fine-tuning, and the testing dataset is pivotal for evaluating the final results [31]. In terms of the training specifics, the Adam optimizer and mean squared error as the loss function were selected. For the direct approach, the model underwent simulation for 100 epochs.

## IV. RESULTS

After conducting simulations for the direct approach, the resulting training and validation loss curves are illustrated in Fig. 6. Initially, the error starts at approximately 1.6 and converges to a training loss of 0.4978 and a validation loss of 0.4519. While the training loss curve displays a smoother progression as expected, the validation loss exhibits more oscillations, possibly due to a less stable model. However, a closer inspection of the actual and prediction plots in Fig. 7 reveals that the prediction curve closely mirrors the trend of the actual curve, indicating high accuracy.

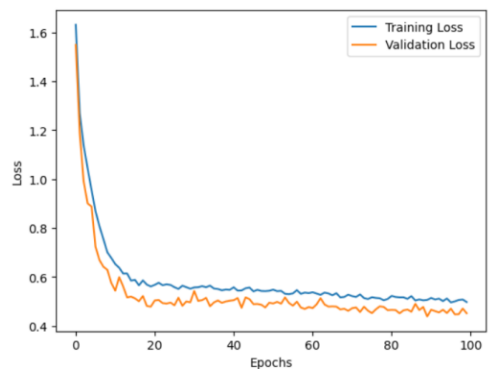

Fig. 6: Training and validation losses for direct approach.

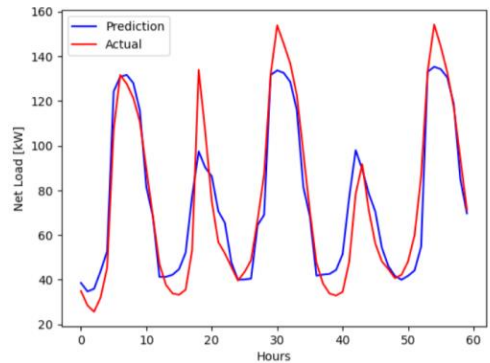

Fig. 7: Actual and predicted net load demand for direct approach.

Upon running simulations for the indirect method, we observed a similar curve where the test loss reached a low of 0.4414, and the validation loss 0.4207 for load demand prediction in Fig. 8. Examining the prediction and actual plots of Fig. 9, it is clear that the prediction aligns closely with the actual trend except at certain peaks.

For wind and solar power predictions in the indirect method, the model achieves a fast convergence in as few as 5 epochs,

providing highly accurate predictions as seen in Fig. 10 and 11. With all values obtained, including load demand, solar power, and wind power, we can compute the net forecasted load using the indirect approach. Fig. 12 demonstrates that the indirect method is slightly better in predicting values, as the prediction curve closely follows the actual curve.

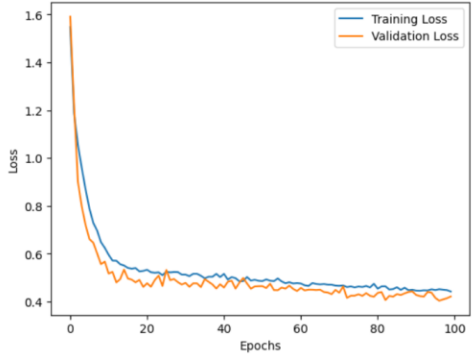

Fig. 8: Training and validation losses for indirect approach.

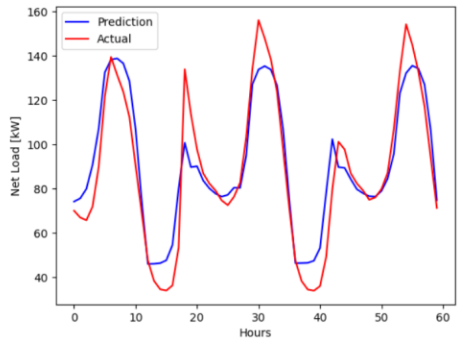

Fig. 9: Actual and predicted load demand of a residential unit for indirect approach.

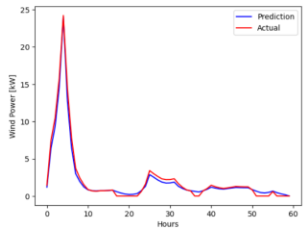 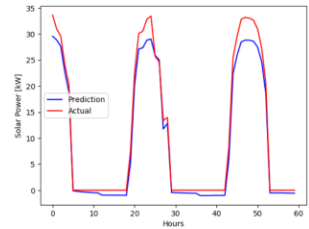

Fig. 10: Actual and predicted wind power for indirect approach

Fig. 11: Actual and predicted solar power for indirect approach

Finally, the histogram of absolute error graph emphasizes that for the indirect method (Fig. 14), accuracy is higher with a maximum absolute error percentage of 50%, while the direct method (Fig. 13) displays errors across the entire 100% range. Moreover, TABLE I presents different errors for both methods, including mean absolute error (MAE), mean squared error (MSE), root mean squared error (RMSE), and normalized root mean squared error (nRMSE) [32], [33]. Notably, the indirect method consistently yields lower values across all metrics.

Anticipating superior performance from the indirect method, the difference is minimal in terms of number, yet the latter method has a much better absolute error distribution with 75% of predictions within a 20% error tolerance compared to 50% for the direct method. The improved performance is due to the establishment of precise relationships between load demand, wind power, and solar power. In conclusion, the indirect method achieves a 9.4% nRMSE, surpassing benchmarks established in various sources in the literature.

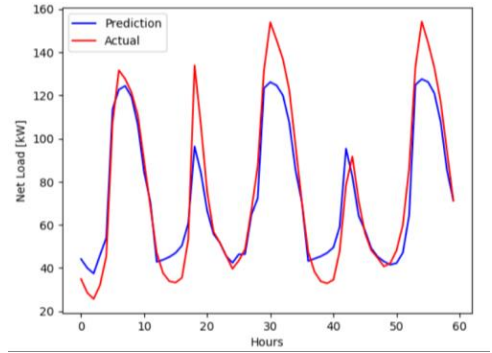

Fig. 12: Actual and predicted net load for indirect approach.

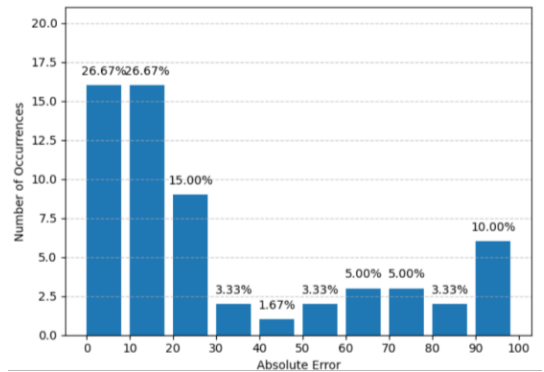

Fig. 133: Histogram absolute error (%) for direct approach (60 Occurrences)

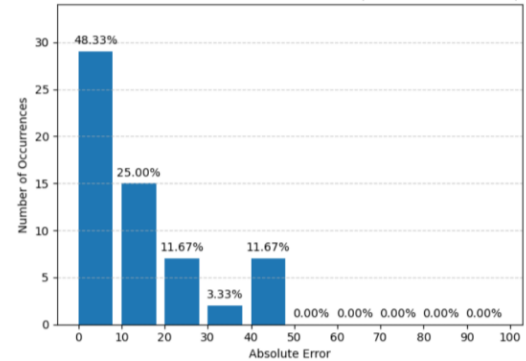

Fig. 144: Histogram absolute error (%) for indirect approach (60 occurrences)

TABLE I. OVERALL COMPARISON OF BOTH NLF METHODS

| Model | Direct Method | Indirect Method |
|---|---|---|
| MAE | 9.48166 | 9.41341 |
| MSE | 153.59356 | 147.63212 |
| RMSE | 12.39329 | 12.15040 |
| nRMSE | 0.09642 | 0.09453 |

V. CONCLUSION

In conclusion, this paper presents a comprehensive investigation into NLF for renewable-based microgrids using a DL model based on LSTM networks. The study considered both solar and wind power, addressing the uncertainty introduced by variable RES. Through a detailed analysis of a residential microgrid test case utilizing a TMY dataset, the proposed LSTM-based model demonstrated remarkable performance in predicting net load.

The comparison between the direct and indirect approaches highlighted the superiority of the indirect method, showcasing a higher accuracy. The indirect approach, involving separate predictions for load demand, wind power, and solar power, proved effective in establishing precise relationships. This approach surpassed benchmarks set by various reference papers, achieving a better loss of 9.4% nRMSE. Additional evaluation metrics such as MAE, MSE, and RMSE, also favored consistently the indirect method, emphasizing its reliability and effectiveness.

The proposed DL model with its LSTM architecture holds promise for enhancing energy management in renewable-based microgrids, providing grid operators with valuable insights for optimal resource utilization.

## References

[1] T. Hong and S. Fan, "Probabilistic electric load forecasting: A tutorial review," *Int. J. Forecasting*, vol. 32, no. 3, pp. 914–938, Jul.–Sep. 2016.

[2] Z. Li, Y. Li, et al., "Deep Learning Based Densely Connected Network for Load Forecasting," *IEEE Transactions on Power Systems*, vol. 36, no. 4, pp. 2829-2840, July 2021.

[3] H. Shi, M. Xu and R. Li, "Deep Learning for Household Load Forecasting—A Novel Pooling Deep RNN," *IEEE Transactions on Smart Grid*, vol. 9, no. 5, pp. 5271-5280, Sept. 2018.

[4] M. Cordeiro-Costas, D. Villanueva, P. Eguía-Oller, M. Martínez-Comesaña, and S. Ramos, "Load Forecasting with Machine Learning and Deep Learning Methods," *Applied Sciences*, vol. 13, no. 13, p. 7933, Jul. 2023.

[5] B. Farsi, M. Amayri, et al., "On Short-Term Load Forecasting Using Machine Learning Techniques and a Novel Parallel Deep LSTM-CNN Approach," *IEEE Access*, vol. 9, pp. 31191-31212, 2021.

[6] J. Yang, M Tuo, J. Lu, and X. Li, "Analysis of Weather and Time Features in Machine Learning-aided ERCOT Load Forecasting", *Texas Power and Energy Conference*, College Station, TX, USA, Feb. 2024.

[7] Kaur, L. Nonnenmacher, C. F.M. Coimbra, "Net load forecasting for high renewable energy penetration grids," *Energy*, vol. 114, pp. 1073-1084, 2016.

[8] G. Tziolis, C. Spanias, et al, "Short-term electric net load forecasting for solar-integrated distribution systems based on Bayesian neural networks and statistical post-processing," *Energy*, vol. 271, 2023.

[9] M. Sun, T. Zhang, Y. Wang, G. Strbac and C. Kang, "Using Bayesian Deep Learning to Capture Uncertainty for Residential Net Load Forecasting," *IEEE Transactions on Power Systems*, vol. 35, no. 1, pp. 188-201, Jan. 2020.

[10] M. Alipour, J. Aghaei, et al., "A novel electrical net-load forecasting model based on deep neural networks and wavelet transform integration," Energy, vol. 205, 2020.

[11] S. Makridakis, E. Spiliotis, V. Assimakopoulos, "Statistical and Machine Learning forecasting methods: Concerns and ways forward," *PLoS ONE*, vol. 13, 2018.

[12] G. Van Houdt, C. Mosquera, G. Nápoles, "A review on the long short-term memory model," *Artificial Intelligence Review*, vol. 53, pp. 5929–5955, 2020.

[13] L. Zhang, N. Gari, L. V. Hmurcik, "Energy management in a microgrid with distributed energy resources," *Energy Conversion and Management*, vol. 78, 2014, pp. 297-305.

[14] U. Kunwar, A. Shupao, et al., "Advantages And Challenges Of DC Standalone Decentralized Microgrid For Institutional Buildings: A Case Study Of The Installed Microgrid In Rewa Engineering College," *2023 3rd International Conference on Intelligent Technologies (CONIT)*, Hubli, India, 2023, pp. 1-5.

[15] M. Grover and O. Egbue, "Analysis of Key Drivers and Challenges Facing Microgrid Deployment," *2020 5th International Conference on Smart and Sustainable Technologies (SpliTech)*, Split, Croatia, 2020, pp. 1-5.

[16] J. Silva-Rodriguez and X. Li, "Water-Energy Co-Optimization for Community-Scale Microgrids," *IEEE 53rd North American Power Symposium*, College Station, TX, USA, Nov. 2021.

[17] C. Zhao, J Silva-Rodriguez, X. Li, "Resilient Operational Planning for Microgrids Against Extreme Events", *Hawaii International Conference on System Sciences*, Maui, Hawaii, USA, Jan. 2023.

[18] National Association of State Energy Officials, "What is a Microgrid?" *Microgrids State Working Group*, [Online]. Available: https://www.naseo.org/issues/electricity/microgrids.

[19] J. Lu, X. Li, et al, "A Synthetic Texas Backbone Power System with Climate-Dependent Spatio-Temporal Correlated Profiles", *arXiv*, Feb. 2023.

[20] NREL, National Renewable Energy Laboratory, "National Solar Radiation Database," [Online]. Available: https://nsrdb.nrel.gov/datasets/archives.html.

[21] Zakariazadeh, S. Jadid, P. Siano, "Smart microgrid energy and reserve scheduling with demand response using stochastic optimization," *International Journal of Electrical Power & Energy Systems*, vol. 63, pp. 523-533, 2014.

[22] R. Chedid, H. Akiki and S. Rahman, "A decision support technique for the design of hybrid solar-wind power systems," *IEEE Transactions on Energy Conversion*, vol. 13, no. 1, pp. 76-83, March 1998.

[23] E. Skoplaki, J.A. Palyvos, "On the temperature dependence of photovoltaic module electrical performance: A review of efficiency/power correlations," *Solar Energy*, vol. 83, issue 5, 2009, pp. 614-624.

[24] Y. Riffonneau, S. Bacha, et al., "Optimal Power Flow Management for Grid Connected PV Systems With Batteries," *IEEE Transactions on Sustainable Energy*, vol. 2, no. 3, pp. 309-320, July 2011.

[25] D. Jones, C.P. Underwood, "A thermal model for photovoltaic systems," *Solar Energy*, vol. 70, issue 4, 2001, pp.349-359.

[26] J. P. Holman, "Heat Transfer," *McGraw Hill Series in Mechanical Engineering*, 10th Edition, 2010.

[27] K. Zuo, "Integrated Forecasting Models Based on LSTM and TCN for Short-Term Electricity Load Forecasting," *2023 9th International Conference on Electrical Engineering, Control and Robotics (EECR)*, Wuhan, China, 2023, pp. 207-211.

[28] S. Ong, N. Clark, "Commercial and Residential Hourly Load Profiles for all TMY3 Locations in the United States," *Open Energy Data Initiative (OEDI)*, 2014, [Online]. Available: https://data.openei.org/submissions/153.

[29] Tesla Photovoltaic Module Datasheet, "T420S, T425S, and T430S," [Online]. Available: https://es-media-prod.s3.amazonaws.com/media/components/panels/spec-sheets/Tesla_Module_Datasheet.pdf.

[30] "Small-Scale Wind," *Energypedia*, [Online]. Available: https://energypedia.info/wiki/Small-Scale_Wind#Village_Power:_Potable_Water.

[31] R. Medar, V. S. Rajpurohit, and B. Rashmi, "Impact of training and testing data splits on accuracy of time series forecasting in machine learning," 2017 *International Conference on Computing, Communication, Control and Automation (ICCUBEA)*, 2017, pp. 1–6.

[32] G. Kavaz and A. Karazor, "Solar Power Forecasting by Machine Learning Methods in a Co-located Wind and Photovoltaic Plant," *2023 12th International Conference on Power Science and Engineering (ICPSE)*, Eskisehir, Turkiye, 2023, pp. 55-59.

[33] L. Li, R. Jing, et al., "Short-Term Power Load Forecasting Based on ICEEMDAN-GRA-SVDE-BiGRU and Error Correction Model," *IEEE Access*, vol. 11, pp. 110060-110074, 2023.